\documentclass[fleqn,usenatbib]{mnras}

\usepackage{newtxtext,newtxmath}

\usepackage[T1]{fontenc}

\DeclareRobustCommand{\VAN}[3]{#2}
\let\VANthebibliography\thebibliography
\def\thebibliography{\DeclareRobustCommand{\VAN}[3]{##3}\VANthebibliography}

\usepackage{graphicx}	
\usepackage{amsmath}	

\title[Microwave response to kink oscillations]{Microwave response to kink oscillations of a plasma slab}

\author[T. I. Kaltman, E. G. Kupriyanova]{
Tatyana I. Kaltman,$^{1}$
and Elena G. Kupriyanova,$^{2}$\thanks{E-mail: elenku@bk.ru (EGK)}
\\
$^{1}$Saint Petersburg Branch, Special Astrophysical Observatory of RAS, Saint Petersburg, 196140, Russia\\
$^{2}$Central Astronomical Observatory at Pulkovo of RAS, Saint Petersburg, 196140,  Russia\\
}

\date{Accepted XXX. Received YYY; in original form ZZZ}

\pubyear{2022}

\begin{document}
\label{firstpage}
\pagerange{\pageref{firstpage}--\pageref{lastpage}}
\maketitle

\begin{abstract}
The modulation of the intensity of microwave emission from a plasma slab caused by a standing linear kink fast magnetoacoustic wave is considered. The slab is stretched along a straight magnetic field, and can represent, for example, a current sheet in a flaring active region in corona of the Sun, or a streamer or pseudostreamer stalk. The plasma density is non-uniform in the perpendicular direction and described by a symmetric Epstein profile. The plasma parameter $\beta$ is taken to be zero, which is a good approximation for solar coronal active regions. The microwave emission is caused by mildly relativistic electrons which occupy a layer within the oscillating slab and radiate via the gyrosynchrotron (GS) mechanism. Light curves of the microwave emission were simulated in the optically thin part of the GS spectrum, and their typical Fourier spectra were analysed. It is shown that the microwave response to a linear kink magnetohydrodynamic wave is non-linear. It is found that, while the microwave light curves at the node oscillate with the same frequency as the frequency of the perturbing kink mode, the frequency of the microwave oscillations at the anti-node is two times higher than the kink oscillation frequency. Gradual transformation the one type of the light curves to another occurs when sliding from the node to the anti-node. This result does not depend on the width of the GS-emitting layer inside the oscillating slab. This finding should be considered in the interpretation of microwave quasi-periodic pulsations in solar and stellar flares.
\end{abstract}


\section{Introduction}
Kink oscillations of solar coronal magnetic structures were discovered in thermal emission in the extreme ultraviolet (EUV) emission (171\,{\AA}) with the Transition Region and Coronal Explorer (TRACE) as a spatial displacements of a coronal loop in the transverse direction \citep{1999ApJ...520..880A, 1999Sci...285..862N}. Since that time, kink oscillations have been one of the most debated magnetohydrodynamic (MHD) mode detected in the corona, especially in the EUV band where the spatial resolution is high enough. For example, the spatial resolution of the Atmospheric Imaging Assembly (AIA) instrument onboard the Solar Dynamic Observatory (SDO) is about 0.7\,Mm (0.6\,arcsec) which is enough to resolve the high-amplitude decaying kink oscillations, with the apparent displacemnt amplitudes ranging in 1--15\,Mm (which is about several minor radii of the oscillating loop) \citep[][]{2019ApJS..241...31N}. Another regime of the kink mode is the low-amplitude decay-less oscillations of coronal loops, first detected by \citet{2012ApJ...751L..27W} 
and \citet{2012ApJ...759..144T}. 
As well as the high-amplitude kink oscillations, the decayless kink oscillations are also eigenmodes of the oscillating loops, with the amplitudes varying from 0.05 to 0.5\,Mm (of the order or even less than the minor radii of the oscillating loops) \citep{2013A&A...560A.107A, 2015A&A...583A.136A}. 
Recently, the higher-amplitude (up to around 1.2~Mm) decay-less kink oscillations were reported by \citet{2021A&A...652L...3M}.
The observational detection of low-amplitude kink oscillations is enhanced by the use of the motion magnification procedure \citep{2016SoPh..291.3251A}. The recently launched Solar Orbiter (SolO) space mission with the on-board Extreme Ultraviolet Imager (EUI) can provide the best available spatial resolution, about 100\,km per pixel \citep[][]{2020A&A...642A...8R} allowing for a more detailed study of kink oscillations \citep[e.\,g.,][]{2022A&A...666L...2M}.

Propagating transverse, kink-like oscillations have also been detected as in coronal jets both in the series of soft X-ray images \citep{2007Sci...318.1580C} and in oscillations of the Doppler shifts in the H\,\texttt{I} Ly$\alpha$ emissions from a jet \citep{2015A&A...573A..33M} as well as in spicules in the H$\alpha$ emission \citep[][]{2014Ap&SS.354..259K}. Propagating kink oscillations are also considered in the coronal streamers at distances from a few to several solar radii \citep[e.\,g.,][]{2010ApJ...714..644C}, and appear as flapping oscillations of the in-situ observed terrestrial magnetospheric plasma sheet  \citep[e.\,g.,][]{2003GeoRL..30.1327S, 2006AnGeo..24.1695P}.

The fine spatial resolution of EUV-imaging instruments allows for distinguishing between the fundamental  and higher spatial harmonics of kink oscillations, based on the harmonic analysis of the light curves \citep[e.g.][]{2004SoPh..223...77V, 2007ApJ...664.1210D, 2016A&A...593A..53P}. Obviously, spatial displacements of a kink-oscillating loop appear as quasi-periodic pulsations (QPPs) of the brightness at the fixed area in the series of images. For example, \citet{2018ApJ...854L...5D} analysed the transverse decay-less oscillations in different segments of a coronal loop, and found QPPs with the fundamental oscillation period and its half. The longer-period component appeared at the loop top, and demonstrated the in-phase behaviour along the whole loop, indicating the fundamental spatial harmonic. The shorter-period oscillations were found to be in anti-phase at the loop legs, and were hence associated with the second harmonic. In another example, both the fundamental and third harmonics of a standing kink mode were detected \citep{2019A&A...632A..64D}. 

Note that the values of the oscillation periods detected in \citet{2018ApJ...854L...5D, 2019A&A...632A..64D} are rather long, ranging from a few to several minutes. 
A statistical study of decaying kink oscillations observed with AIA revealed that their periods vary from 1 to 28 minutes \citep[][]{2019ApJS..241...31N}. Indication of a shorter oscillation period, about 45~s was found, for example, by \citet[][]{2022FrASS...932099L}. 
For the shorter periods, it becomes a problem to detect them and their higher harmonics in the EUV data because of insufficient temporal resolution. For example, the SDO/AIA data have the time binning of 12\,s. Recently, it became possible to search for shorter period kink oscillations with SolO/EUI which can take images with the  time cadence of a few seconds \citep{2022MNRAS.516.5989Z}. 
Using these data, \citet{2022arXiv220505319P} have found a short loop oscillating in the transverse direction with the period around 14\,s, in the decay-less mode.

Light curves of the solar coronal radio emission have much higher time resolution, up to 10--100\,ms. However, the intrinsically low spatial resolution does not allow one to detect  kink displacements of coronal plasma structures directly from the images and, therefore, forces one to use indirect methods of the identification of kink oscillations. The microwave emission offers an important and unique information about parameters of the accelerated particles, the magnetic field, and other macroscopic parameters of the emitting plasma. In particular, the microwave emission is sensitive to the MHD plasma parameters which are perturbed by coronal MHD waves, allowing for their detection \citep{1983SoPh...87..177T}. 

As the spatial resolution of available radio instruments is insufficient for the direct detection of kink displacements in the corona, there is a need for indirect methods. One of them is connected with the appearance in the time signal multiple harmonics of the leading periodicity, which could be found with the use of the  Fourier transform \citep[e.\,g.,][]{2009A&A...493..259I}, wavelet transform \citep[e.\,g.,][]{2013SoPh..284..559K}, and the Empirical Mode Decomposition method \citep[e.\,g.,][]{2015A&A...574A..53K}. The ratios of oscillation periods, associated with different parallel harmonics of the standing kink mode may carry specific signatures of the kink mode \citep[e.g.,][]{2009SSRv..149....3A}.

For example, spatial information has been combined with the detection in the time domain of an oscillatory variation of the microwave flux with 31\,s and 21\,s, to reveal a kink oscillation \citep{2013SoPh..284..559K}. The longer period oscillations were sitated near the loop top, i.\,e., near the anti-node of the fundamental harmonic of the kink mode, where the  amplitude is maximal. The shorter period oscillations were detected at the loop legs, i.\,e., near the anti-nodes of the second harmonic of the standing kink oscillation. Note that this result was obtained using data of the Nobeyama Radioheliograph at 17\,GHz, with the beam size is about 12\,arcsec and the pixel size of 2.5\,arcsec \citep[][]{1994IEEEP..82..705N}, which did not allow the direct detection of the kink-oscillating loop. 

Another problem in the detection of the kink oscillations is their rapid damping \citep[in the large amplitude decaying regime,][]{2021SSRv..217...73N}, which significantly shortens the duration of the oscillatory signal \citep[][]{2019ApJS..241...31N}. In particular, it broadens the corresponding spectral peaks, reducing its amplitude. Therefore, an independent diagnostics of the kink mode, accounting for the short lifetime of the kink mode-associated QPPs, would be highly useful.

It should be accented here that oscillation periods of the kink mode are usually taken to be equal to the estimated periods of QPPs, within certain error bars \citep[e.\,g.,][ etc.]{2003ARep...47..873Z, 2013SoPh..284..559K, 2022MNRAS.511.2880S}. However,
in the microwave band, such the direct association is risky considering the non-linear dependence of the intensity of the microwave emission on the parameters of the emitting source, and, in particular, parameters perturbed by a kink oscillation.

The microwave emission produced by solar flares is generally associated with the non-thermal gyrosynchrotron (GS)
emission of mildly relativistic electrons accelerated during the flare. 
The intensity of the emission depends in a complex way on many parameters of both the background plasma and accelerated electrons \citep[][]{1965ARA&A...3..297G, 1968Ap&SS...2..171M}. In particular, the GS emission  depends non-linearly on the magnetic field, including its magnitude and the direction with respect to the line-of-sight. The latter parameter is intrinsically perturbed by a kink wave of a finite wavelength \citep{2003A&A...397..765C}. Moreover, in the optically thin regime, the kink wave causes the variation of the instantaneous column depth of the emitting segment of the waveguide. 
In the modelling of the modulation of the GS emission by an MHD wave, it also also necessary to take into account that the GS-emitting electrons do not necessarily occupy the whole oscillating volume \citep[][]{2022MNRAS.516.2292K}. Besides, filamentation of a coronal loop, caused by the development of the Kelvin-Helmholtz instability \citep[e.\,g.,][]{2019FrP.....7...85A}, could strongly affect the modulation depth of its microwave emission \citep{2022ApJ...937L..25S}.

The aim of the paper is to study the microwave response to a kink oscillation of a plasma slab. The GS-emitting electrons 
do not occupy the whole
oscillating volume. The model of the plasma slab, the source of the GS emission, the kink wave, and the procedure of simulations of the microwave fluxes are described in Section~\ref{s:Model}. Results are summarized in Section~\ref{s:results}, and discussed in Section~\ref{s:Discussion}. Conclusion is given in Section~\ref{s:Conclusions}.

\section{Model}\label{s:Model}
\subsection{Plasma slab with GS source}\label{s:Slab}
\begin{figure*}
	\centering{
		\includegraphics[width=1.0\textwidth]{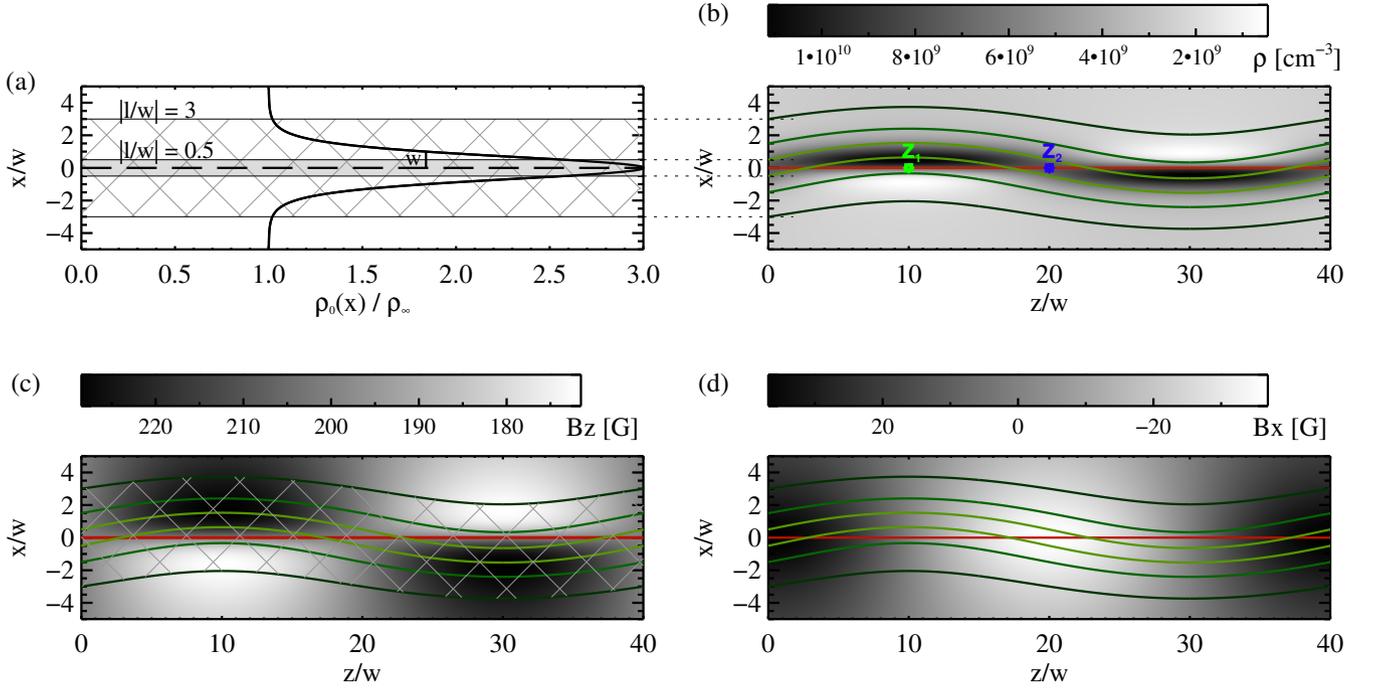}
	}
	\caption{ Panel (a): normalised transverse profile of the unperturbed thermal plasma density. The vertical short line indicates the characteristic  half-width $w$. Two rectangles illustrate homogeneous distributions of the density of non-thermal electrons corresponding to a narrow GS source ($|l/w| = 0.5$, grey area) or a wide GS source ($|l/w| = 3$, cross-hatched area). Panel (b): the perturbed plasma density $\rho$. Panel (c): the perturbed  $B_{z}$ component of the magnetic field. Panel (d): the perturbed  $B_{x}$ component of the magnetic field. The darkest colour corresponds to the maximal value, while the lightest colour~--- to the minimal value. The green point $z_1$ in panel (b) denotes one of the kink wave nodes and the blue point $z_2$~--- one of the anti-nodes. Different pairs of the same-coloured wavy lines delimit the GS sources having different widths. An example of the perturbed wide GS source is as a cross-hatched area in panel (c). The slab axis is marked by the long-dashed line in panel (a) and with the red line in panels (b), (c), (d).}
	\label{f:rho_Bx_Bz_kink}
\end{figure*}
We use the plasma slab model introduced by \citet{2003A&A...409..325C} and described in detail in our previous paper \citep{2022MNRAS.516.2292K}. Therefore, here, we discuss it briefly. We consider a plasma slab which is stretched along the uniform magnetic field $B_0$ directed along $z$-axis ($B_{z0} = B_0 = \textrm{const}$, $B_{x0} = 0$, $B_{y0} = 0$). The plasma density is non-uniform in the perpendicular direction, along the $x$-axis, and uniform in the $y$ direction. The slab is observed by an infinitely remote observer with the line-of-sight in the $xz$ plane. Thus, we may ignore the $y$ axis and perform MHD modelling in the the $xz$ plane. 

The transverse profile of the plasma density is described by the symmetric Epstein function \citep[e.g.][]{1995SoPh..159..399N, 2003A&A...409..325C},
\begin{equation}\label{eq:Epstein}
	\rho_0(x) = \rho_\mathrm{max} \rm{sech}^2 \it \left( \frac{x}{w} \right)+ \rho_{\infty},
\end{equation}
where $\rho_\mathrm{\max}+\rho_{\infty}$, $\rho_{\infty}$ and $w$, are, respectively, the density at the centre and infinity, and the half-width of the slab, see Figure~\ref{f:rho_Bx_Bz_kink}. The density contrast is defined as
\begin{equation}\label{eq:dc}
	d=\frac{\rho_\mathrm{\max}+\rho_{\infty}}{\rho_{\infty}}.
\end{equation}
In the following, we consider a zero-$\beta$ plasma typical for the corona. In this limit, the total pressure balance requires the equilibrium magnetic field $B_0$ to be constant everywhere.

A hot plasma within the slab emits via the thermal free-free mechanism. Additionally, a portion of co-existing non-thermal, accelerated electrons, gyrating around the magnetic field, produces the GS emission, which is orders of magnitude stronger than the free-free emission. Under typical conditions in the solar corona both types of the emission are observed in the microwave band. We consider the thermal plasma temperature $T = 10^6$\,K, the magnetic field $B_0 = 200$\,G, the number density at the slab axis $\rho_0(x=0) = 5 \times 10^9$\,cm$^{-3}$ and the number density infinitely far from the slab $\rho_\infty = \rho_0(x=0)/(d-1)$\,cm$^{-3}$. Here, $d=3$ is chosen.
For simplicity, we consider the accelerated electrons isotropically distributed over the pitch-angle and power-low distributed over energies from 0.1 to 10\,MeV with the spectral index $\delta = 3$. The number density of the non-thermal electrons is chosen to be $n_1 = 5 \times 10^5$\,cm$^{-3}$. This value is constant in the unperturbed GS-emitting source.

According to aim of this study, we consider the system, where the accelerated electrons fill only a part of the slab~--- the area between two magnetic field lines at the distance $x/w=|l/w|$ from the slab axis. Here,  $|l/w|$ means the half-width of the GS-emitting source. Hereinafter, all the spatial quantities are normalized over the characteristic slab width $w$. In the same manner as in \citep[][]{2022MNRAS.516.2292K}, we vary $|l/w|$  from 0.1 to 3, i. e., the GS sources could be both narrower ($|l/w| \leq 1$) and wider ($|l/w| > 1$) than the characteristic slab width. Panel (a) of Figure~\ref{f:rho_Bx_Bz_kink} shows examples of two different widths: the grey rectangle corresponds to the narrow GS source ($|l/w| = 0.5$) and the cross-hatched rectangle~--- to the wide GS source ($|l/w| = 3$). Below, we omit the modulus sign, for simplicity.

\subsection{The kink wave}\label{s:Kink}
The slab is perturbed by a kink fast magnetoacoustic MHD mode. The perturbing wave is considered to be plain in the $y$ direction, and hence is not subject to  the effect of resonant absorption. Kink perturbations are described by ideal MHD equations linearised near the considered equilibrium \citep[see, e.g.,][]{1997JPlPh..58..315N} in a zero-$\beta$ plasma. The perturbed plasma density and magnetic field components can be expressed via the transverse component of the perturbed speed in the $x$ direction, $\tilde{V}_x(x,z,t)$ as
\begin{equation}\label{eq:rho_B}
\begin{aligned}
& \tilde{\rho} = - \int \frac{\partial (\rho_0 \tilde{V}_{x})}{\partial x} dt, \\
	& \tilde{B}_{x} = B_0 \int \frac{\partial \tilde{V}_{x}}{\partial z} dt, \:\:\:\:\:\:\:\:\:\:\:\:
	\tilde{B}_{z} = - B_0 \int \frac{\partial \tilde{V}_{x}}{\partial x} dt. 
\end{aligned}
\end{equation}
The transverse component $\tilde{V}_{x}$ of a standing kink wave with the wavelength is $\lambda = 2\pi/k$, has the form
\begin{equation}\label{eq:Vx}
	\tilde{V}_{x}(x,z,t) = A U(x) \sin(kz)\cos(k V_\mathrm{ph} t),
\end{equation}
where $A$ is the amplitude normalized to the Alfv\'en speed $C_\mathrm{A0}$ at the slab axis ($x=0$), $A \ll 1$, $V_\mathrm{ph} = \omega / k$ is the phase speed, and $k$ is the parallel wave number. The use of the Epstein profile (Equation~(\ref{eq:Epstein})) allows us to obtain the expression for the  structure $U(x)$ of the kink wave in the analytical form  \citep{2003A&A...409..325C},
\begin{equation}\nonumber
	U(x) = \mathrm{sech}^{\mu}(x/w), \:\:\:\:\:\:\:\:\:
	\mu = \frac{ |k|w }{ C_\mathrm{A \infty }} \sqrt{ C_\mathrm{A \infty}^2 - V_\mathrm{ph}^2},
\end{equation}
where $C_\mathrm{A \infty}$ is the Alfv{\'e}n speed at the infinitely remote point ($x \to \infty$). The phase speed $V_\mathrm{ph}$ is defined by the dispersion equation 
\begin{equation}\nonumber
    \sqrt{C_\mathrm{A \infty}^2 - V_\mathrm{ph}^2} = |k|w \frac{C_\mathrm{A \infty}}{C_\mathrm{A0}^2} \left(V_\mathrm{ph}^2 - C_\mathrm{A0}^2\right).
\end{equation}
The dispersion relation can be rewritten as a bi-quadratic equation, which gives us an exact analytical solution. 
Considering Equations~(\ref{eq:rho_B}) and the initial conditions $B_{z0} = B_0 = \textrm{const}$ and $B_{x0} = 0$, the perturbed values for plasma density and the magnetic field components are
\begin{equation}
	\rho = \rho_0 + \tilde{\rho}, \:\:\:\:\:\:\:\:\:  
	B_{z} = B_0 + \tilde{B}_{z}, \:\:\:\:\:\:\:\:\:  
	B_{x} = \tilde{B}_{x}.
	\label{eq:RhoB_Total}
\end{equation}
Panels (b), (c), (d) in Figure~\ref{f:rho_Bx_Bz_kink} illustrate snapshots of these quantities (by the gradient of grey) in the $xz$ plane at the instant of time when the perpendicular velocity amplitude is maximal. The wavelength was chosen to be
$\lambda=40w$ and its relative amplitude~--- to be $A = 0.05$. We consider $w = 3 \times 10^8$\,cm. 
It was taken that the number density of the accelerated electrons vary in the inverse proportion with the local distance between magnetic field lines limiting the GS source.
Since the kink mode is a low-compressive mode, this distance and, accordingly, the number density of the non-thermal electrons vary within no more than 12\%, in our model.

We consider the slab inclined relatively to the line-of-sight by the viewing angle $\theta$. Here, $\theta$ is angle between the $z$ axis of the unperturbed slab and the line-of-sight.

\subsection{Calculations of microwave emission}\label{s:microwave}
The kink wave perturbs the plasma density and the magnetic field, as well as the number density of the accelerated electrons at each pixel of the plasma slab. The modulation of the slab parameters leads to the modulation of the microwave emission.  In our study, to calculate the microwave emission, we apply the \textsc{Fast Gyrosynchrothon Codes} \citep[FGS codes,][]{2021ApJ...922..103K} where the exact expressions for both emission and absorption coefficients were used (see Equations~(1a) and (1b) there). The codes obtain these coefficients at each voxel of the slab accounting for the both thermal free-free and GS mechanisms. The voxel size of $0.02w \times 0.02w \times 10w$ was chosen along the $x$, $z$, and $y$ directions, respectively. Then the radiative transfer equation is solved along the line-of-sight providing as the result both the left-hand and the right-hand polarized components of the emitted microwave at each pixel of the plane-of-the-sky. We calculate the total emission intensity as the sum of both polarized components. Varying the viewing angles $\theta$ within the interval $50^\circ \leq \theta \leq 89^\circ$ we obtain the intensity of the microwave emission in the plane-of-the-sky projection.

To understand clearly how the kink mode affects the microwave radiation in the different parts of the perturbed slab we draw lines-of-sight through some selected points at the slab axis $z$. To simulate a light curve at a selected point, we apply sequential calculations for several phases of the full cycle of the MHD wave, using the corresponding parameters of the background plasma and accelerated electrons described in Section~\ref{s:Kink}. We consider twenty such phases, regularly distributed within one oscillation cycle. The period of the kink wave equals to $P_\textrm{kink} = 20$~time units (t.\,u.). 

Additionally, to study how the character of the light curve depends on the exact location along the $z$ axis 
we consider in detail the interval from $z_1$ (the kink wave anti-node) to $z_2$ (node) (see panel (b) in Figure~\ref{f:rho_Bx_Bz_kink}). We divide the interval by ten equal sub-intervals (so, each interval has the length equal to $w$) and simulate the light curves for the plane-of-the-sky projections of $z_1$, $z_2$, and nine positions between them. 

Note that we analyze here the emission at 17\,GHz where the emission is optically thin for each the considered width of the GS source.

\section{Results}\label{s:results}
\begin{figure*}
	\centering{
		\includegraphics[width=0.90\textwidth]{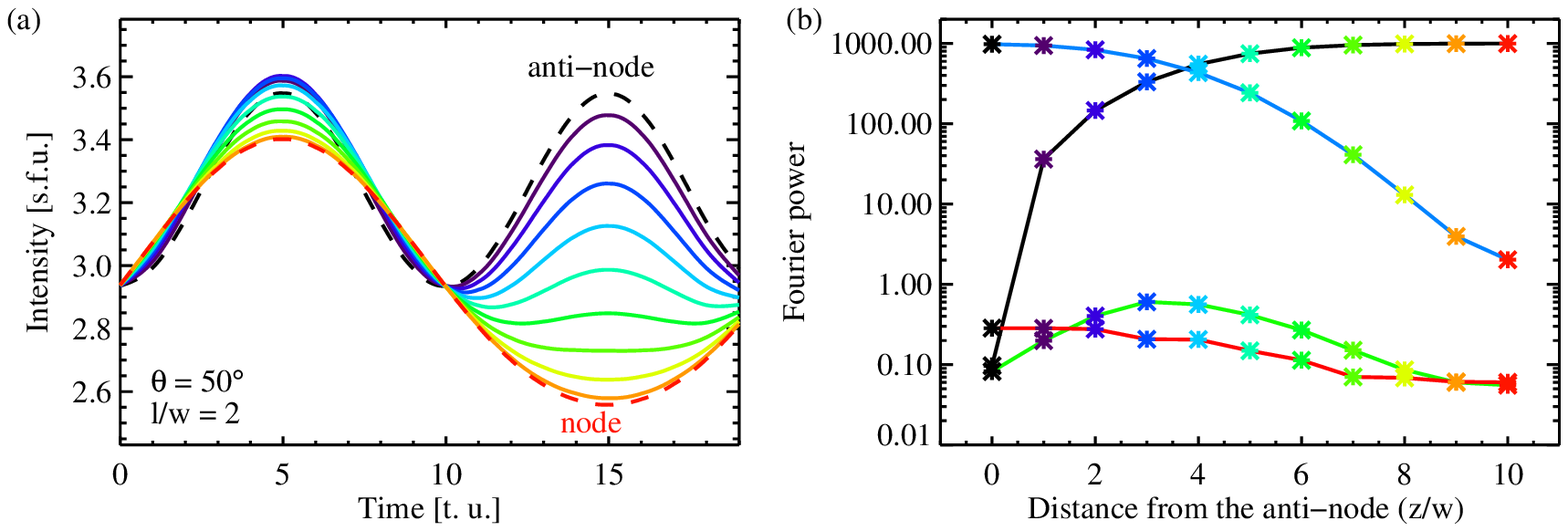}
		\includegraphics[width=0.90\textwidth]{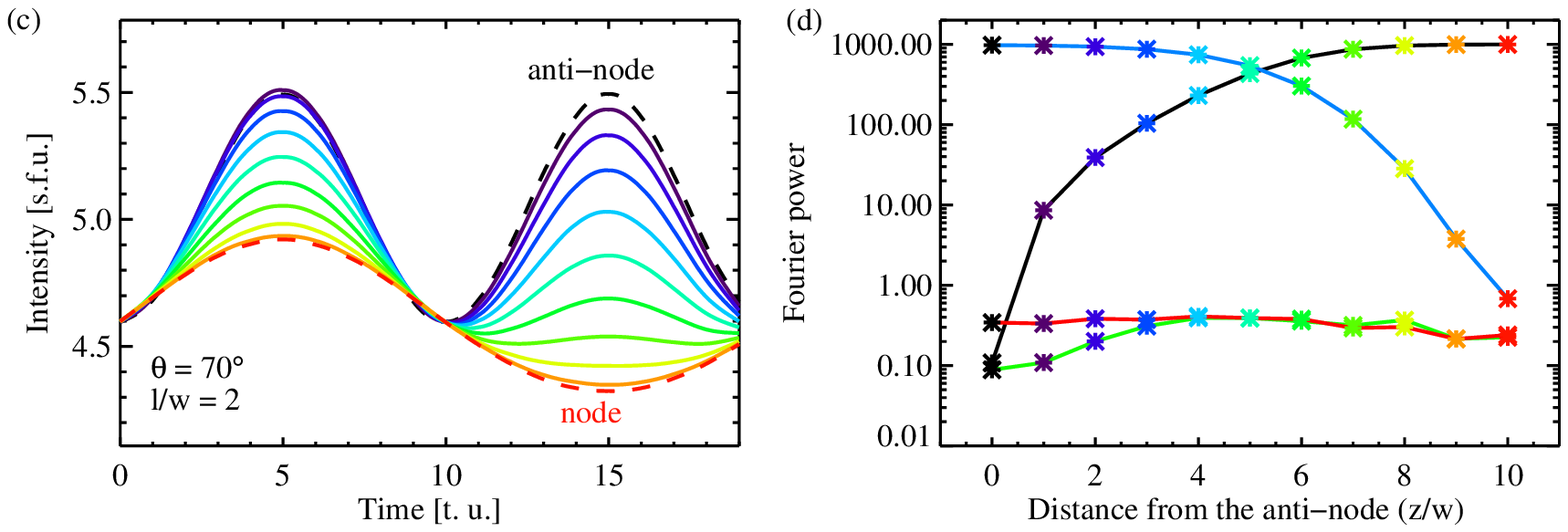}
		\includegraphics[width=0.90\textwidth]{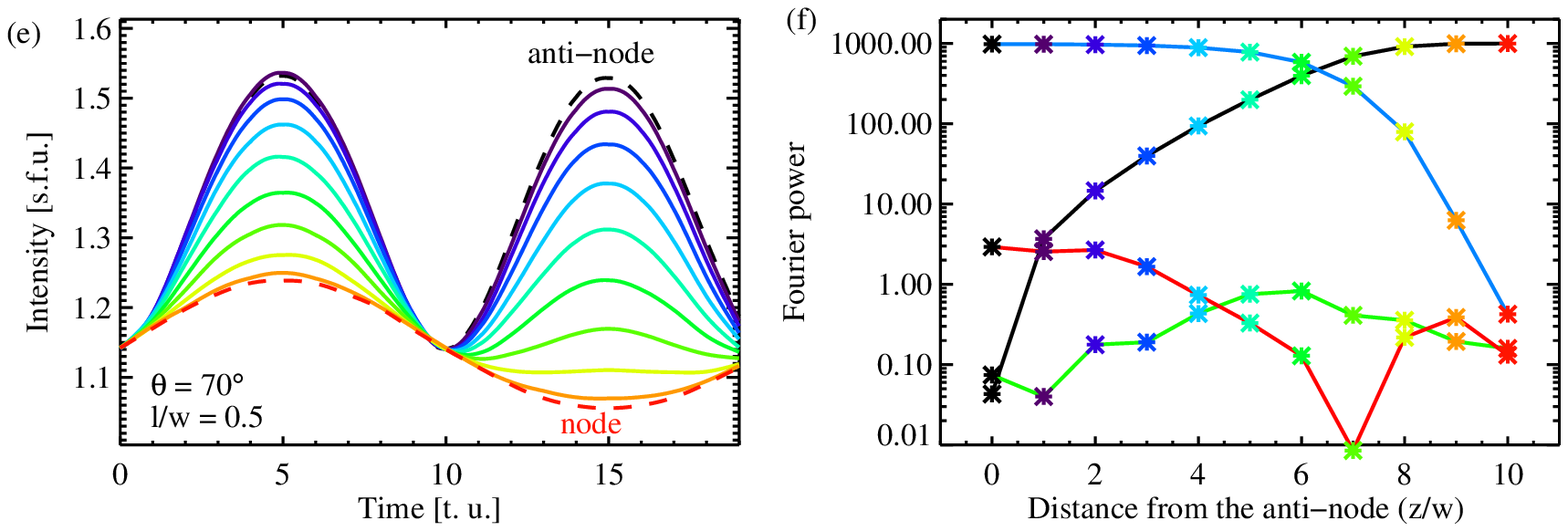}}
	\caption{Left panels: microwave light curves at $f = 17$\,GHz from a kink-oscillating slab at different MHD wave segments: form the anti-node (the black dashed curve, $z_1$ point in panel (b) in Figure~\ref{f:rho_Bx_Bz_kink}) to the node (the red dashed curve, $z_2$ point). The right panels: the corresponding Fourier powers (asterisks) of the fundamental harmonic in the Fourier periodogram (black curve, the period is 20\,t.\,u.) second harmonic (cyan curve, the period is 10\,t.\,u.), third harmonic (green curve, the period is 6.67\,t.\,u.), and fourth harmonic (the period is 5\,t.\,u.). The colour of each family of the asterisks (at the fixed $z/w$) corresponds to the colour of the light curve in the left  panel. Panels (a) and (b) correspond to the wide GS source ($l/w = 2$) inclined to an observer by the viewing angle $\theta = 50^\circ$, panels (c) and (d)~--- to $l/w = 2$ and $\theta = 70^\circ$, panels (e) and (f)~--- to the narrow GS source, $l/w = 0.5$, and $\theta = 70^\circ$.}
	\label{f:many_z}
\end{figure*}
Examples of obtained light curves, interpolated to a smaller time bin, are shown in Figure~\ref{f:many_z} (the left  panels). In this figure, different colours indicate different locations between $z_1$ and $z_2$, inclusively. The black dashed curve corresponds to the kink wave anti-node ($z_1$ point in Figure~\ref{f:rho_Bx_Bz_kink}, panel (b)). The red dashed curve relates to the kink wave node ($z_2$ point in the same panel). Nine intermediate colours correspond to nine intermediate locations between the anti-node and node. This figure illustrates three cases: panel (a) corresponds to the wide GS source ($l/w = 2$) with the unperturbed slab inclined to the observer by the viewing angle $\theta = 50^\circ$, panel (c)~--- $l/w = 2$ and $\theta = 70^\circ$, panel (e)~--- to the narrow GS source, $l/w = 0.5$, and $\theta = 70^\circ$.

As we can see in Figure~\ref{f:many_z}, during the MHD oscillation cycle, the light curves for the node and the nearby points show a corresponding periodic variation. Then, passing from the node toward the anti-node, during the second half of the oscillation period (time interval from 10 to 20\,t.\,u.) the light curve first becomes flatter, then an extra peak arises instead of the minimum, and then the amplitude of this secondary peak increases, reaching a maximum at the anti-node. Thus, while at the node the microwave oscillations occur with the period of the kink wave, the oscillation period is twice shorter at the anti-node. 

Another visualisation of the obtained result is presented in the right  panels of Figure~\ref{f:many_z} which shows the corresponding Fourier powers of the microwave emissions coming at different distances $z/w$ from the anti-node. The values $W_1$ of the fundamental Fourier harmonic are plotted by the black curve (the period is 20\,t.\,u.), while $W_2$ for the second harmonic~--- by the cyan curve (the period is 10\,t.\,u.). We add here the Fourier powers of the third harmonic (the green curve, the period is $P_1/3 \approx 6.67$\,t.\,u.) and of the fourth harmonic (the period is $P_1/3 = 5$\,t.\,u.) as they also appear in the periodogram as a result of the anharmonicity of the obtained microwave light curves. The colour of each family of the asterisks (at the fixed $z/w$) corresponds to the colour of the light curve in the left  panel. 

Figure~\ref{f:many_z} shows that, for each considered combinations of the parameters $l/w$ and $\theta$, either the fundamental or second Fourier harmonics dominates over the others. The Fourier power $W_1$ decreases and, simultaneously, the Fourier power $W_2$ increases when passing from the node to anti-node. Moreover, Figure~\ref{f:many_z} reveals similar properties of the light curves: $W_1 > W_2$ at least at three positions around the node (red to green colours) while $W_2 > W_1$  at least at three positions around the anti-node (black to blue colours). Around the midpoint between $z_1$ and $z_2$ (light-blue and blue-green colours), the domination of either the fundamental or second harmonic depends on the exact combination of $l/w$ and $\theta$.

\begin{figure*}
	\centering{
	\includegraphics[width=0.40\textwidth]{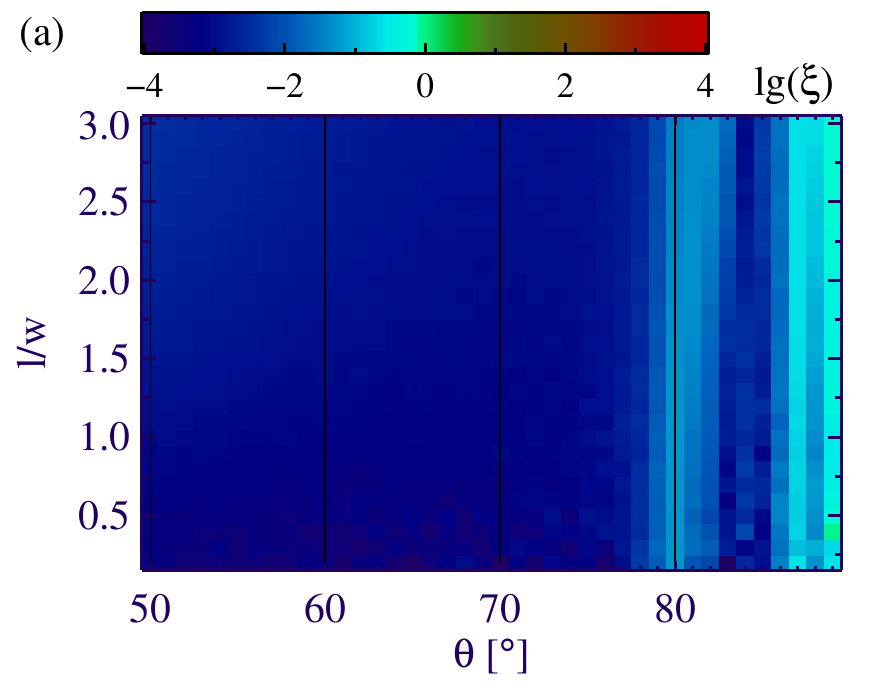}
    \includegraphics[width=0.40\textwidth]{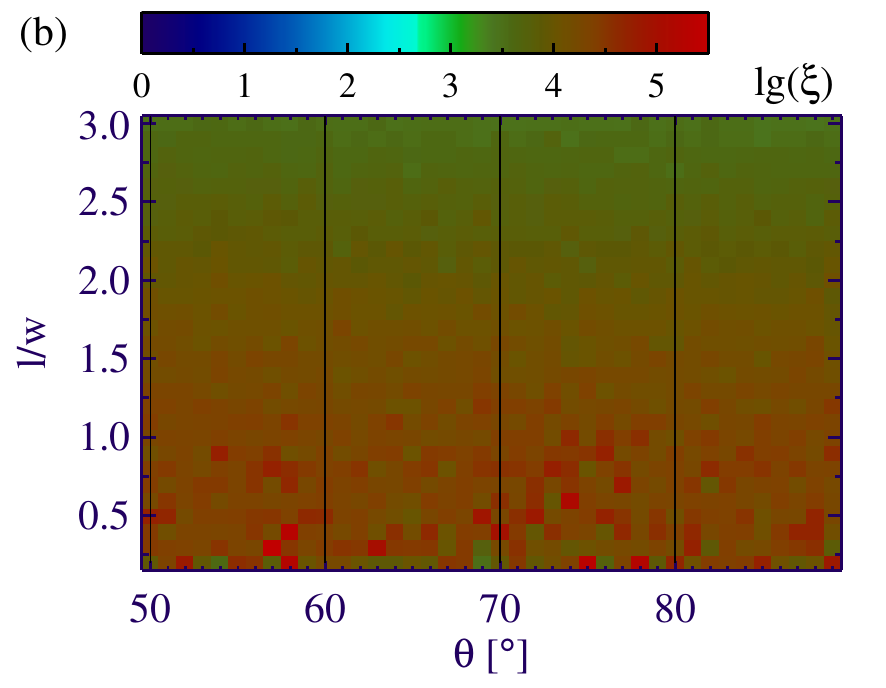}}
	\caption{Distributions of the degree of non-linearity ($\log_{10} \xi$) relatively to the width of the GS source ($l/w$) and the viewing angles $\theta$, in the optically thin regime at 17\,GHz. 
	Panel (a) corresponds to the kink wave node, while panel (b)~--- to the anti-node.}
	\label{f:xi_kink}
\end{figure*}

In order to illustrate how this effect depends on both the width of the GS source ($l/w$) and the viewing angle ($\theta$), we introduce  parameter~--- the degree of the non-linearity, $\xi = W_2/W_1$ \citep[see its first application in][]{2022MNRAS.516.2292K}. If $\log_{10}(\xi) < 0$ then the non-linearity is weak while if $\log_{10}(\xi) \geq 0$ then the non-linearity is strong. The results are shown in Figure~\ref{f:xi_kink} for the node (panel(a)) and the anti-node (panel (b)). 
From this figure, it is evident that the degree of non-linearity keeps almost  the same value for any combination of the parameters $l/w$ and $\theta$ for both the node (weak non-linearity) and the anti-node (strong non-linearity). The exception is some area between $79^\circ$ and $89^\circ$ for the node (in the left panel) where the  non-linearity is stronger than for $\theta < 79^\circ$. This peculiarity is because the local angle $\theta$  passes through 90$^\circ$, within a segment (or segments) of the line-of-sight where the conversion between the ordinary and extraordinary modes occurs  due to the quasi-transverse propagation effect.

\section{Discussion}\label{s:Discussion}
The principal result of this study is the demonstration that the modulation of a microwave signal by a linear kink wave can be highly non-linear. A similar result has been obtained for an essentially compressive sausage mode \citep[][]{2022MNRAS.516.2292K} in a simplified 2D slab model, and in a 3D loop with a fine structure \citep[][]{2022ApJ...937L..25S}. This non-linearity is totally caused by the gyrosynchrotron mechanism of emission.
The character of the non-linearity for the kink mode essentially differs from that of the sausage modes. For the sausage mode, the dependence of the apparent nonlinearity of the microwave signal on both the width of the GS source and viewing angle was found to be significant. For the kink mode, the dependence of the $\xi$ value on these parameters is very weak.

In contrast, for the kink mode, another kind of the non-linearity was detected: the strong dependence of the degree of the non-linearity on the location of the emitting source with respect to the kink oscillation phase (nodes or anti-nodes in the case of a standing wave). This result opens a interesting prospective for diagnostics of kink waves in the microwave band. 

On one hand, this result allows identifying the kink mode in the low-quality QPPs based on the analysis of the light curves from different segments of the oscillating structure.
On the other hand, the result that the spectral properties of the time profiles of the microwave signal are keeping within finite vicinities ($\leq |3 w|$) of the node or anti-node has an observational applicability. In the current study, all the spatial scales are normalized by $w$. Therefore, if we set up $w = 3$\,Mm, the vicinity (in one direction from the node or anti-node) will be 9\,Mm, which corresponds to around 12\,arcsec. Increasing $w$ by a factor of two, makes the vicinity two times wider.
This size is comparable with the beam sizes of the current and future instruments in the microwave band. Among them, the  Radio Telescope of Russian Academy of Sciences~--- 600  \citep[RATAN-600,][]{2011AstBu..66..205B}, the Siberian Radioheliograph \citep[SRH,][]{2020STP.....6b..30A}, the  Expanded Owens Valley Solar Array \citep[EOVSA, ][]{2018ApJ...863...83G},  the Mingantu Spectral Radioheliograph \citep[MUSER,][]{2009EM&P..104...97Y}, etc. 
The spatial resolution is proportional to the ratio of the wavelength of observations to the aperture of the telescope. Obviously, for other parameters being fixed, the better spatial resolution is provided at the higher frequencies. The highest frequency bands of the mentioned instruments fall into the optically thin part of the gyrosynchrotron spectrum, which is the case studied in our paper.

At the moment, the best spatial resolution is reached at the Karl G. Jansky Very Large Array (VLA), where the  beam size is 0.2 arcsec to 0.04 arcsec. However, note that VLA  observes the Sun \citep[e.\,g.,][]{2021AGUFMSH24B..05L} but too rarely. The spatial resolutions of the instruments which observed the Sun routinely (current, archive data, or future) are 1.3\,arcsec at 15\,GHz for MUSER, 5\,arcsec resolution at 34\,GHz for NoRH, 6\,arcsec at 18\,GHz for EOVSA, and 7--13\,arcsec at 12--24\,GHz for SRH. Besides, the one-dimensional scans are obtained by RATAN-600 which has a knife-edge 18\,arcsec by 15\,arcmin beam at 15\,GHz.
The start of a new observation complex 
with the implemented tracking mode will allow to RATAN-600 receiving a signal from a selected active region during several tens of minutes in more than 8000 frequency channels/GHz and with time resolution up to 8\,ms/spectrum \citep[][]{2020gbar.conf..407S}. We should also mention the future Square Kilometre Array \citep[SKA,][]{2015aska.confE.169N} which is expected to provide the record spatial resolution because of its giant aperture. 

The high time resolution and high sensitivity of the radio instruments, in combination with the spatial resolution of the archive, current, upgraded, and the future instruments, will allow microwave observations of different segments of kink-oscillating plasma structures to become feasible.


\section{Conclusion}\label{s:Conclusions}
We studied the modulation of the intensity of microwave emission from a plasma slab caused by a linear standing  kink wave. The accelerated electrons, which are the source of the gyrosynchrotron emission, occupy a layer within the oscillating slab. Light curves of the microwave emission were simulated in the optically thin part of the GS spectrum. It is shown that the microwave response on the linear kink wave in anharmonic (non-linear), and this non-linearity is totally caused by the gyrosynchrotron mechanism. We found that the microwave light curves of the emission from the kink wave node oscillate with the same frequency as the frequency of the perturbing kink mode. In contrast, the frequency of the microwave oscillations at the anti-node is twice higher than one of the kink mode. It was shown that gradual transformation the one type of the light curves to another occurred when sliding from the node to the anti-node. This result does not depend on the width of the GS-emitting layer. The characteristic anharmonic microwave signals coming from some vicinity of the node (or anti-node) are similar at least within three widths of the slab in both directions. This distance is comparable with the beam size of existing instruments. Thus, we may expect that the microwave emission integrated in the beam would have the same time signature. It would allowing for the detection of the kink oscillation modulation signal, provided the nodes and anti-nodes are spatially separated by more than several widths of the slab.  

If considering a diagnostic potential of the results obtained in this study we should accent on the ratio of the periods of higher harmonics detected in the Fourier power spectrum of the light curve to the fundamental period ($P_1 = P_\mathrm{kink}$). In our study, we found that the ratio should equals to $P_n/P_1 \approx 1/n$, where $n$ is the harmonics number, $n> 1$. This ratio is attributed to the nonlinear nature of the microwave emission mechanism. On the other hand, if the observed signal consists of different spatial harmonics of an MHD oscillation, the ratio is expected to be $P_{\mathrm{kink}_n}/P_\mathrm{kink} > 1/n$ because of the wave dispersion \citep[e.\,g.,][]{2009A&A...493..259I} or stratification \citep[e.g.][]{2009SSRv..149....3A}. This feature should be taken into account in the identification of the higher spatial harmonics in observational data.

It should be noted that we considered here the simplest oscillating and emitting system which allowed us to demonstrate, for the first time, the pure line-of-sight effect on the microwave emission. 
Particularly, the symmetric Epstein profile was chosen as, in contrast with all other known profiles, it gives an explicit solution to the eigenvalue problem, i.\,e., in this case we do not need to solve the dispersion relation numerically. The effect of the departure from this assumed profile on the microwave signature of kink oscillation is of interest: even small changes in the magnetic field distribution within the slab could lead to detectable changes in the microwave intensity. 
Definitely, more realistic models should be considered accounting for other transverse density profiles \citep[e.g.,][]{2021MNRAS.505.3505K, 2007A&A...475..341V, 2002ApJ...577..475R},  as well as the 3D geometry, the wave dispersion including the fast kink wave trains \citep[e.\,g.,][]{2021MNRAS.505.3505K, 2022MNRAS.515.4055G}, and, very importantly, the convolution with the beam size of a radio instrument.

\section*{Acknowledgements}

This study is supported  under the
Ministry of Science and Higher Education of the Russian Federation grant 075-15-2022-262 (13.MNPMU.21.0003). 

\section*{Data Availability}

The results obtained in the paper are theoretical; no real observations have been used. Distributions of parameters in the slab were obtained using equations in Section~\ref{s:Model}, and their digital version is available on request to the corresponding author.



\bsp	
\label{lastpage}
\end{document}